\documentstyle[11pt]{article}\textheight 230mm\textwidth 150mm
            \newcommand{\be}{\begin{eqnarray}}
            \newcommand{\ee}{\end{eqnarray}}
           \newcommand{\eel}[1]{\label{#1}\end{eqnarray}}
\newcommand{\e}[1]{\label{e:#1}\end{eqnarray}}
     
            \newcommand{\ie}{{\em i.e.\ }}

            \newcommand{\la}{{\lambda}}
            
            \newcommand{\del}{{\delta}}
 \newcommand{\om}{{\omega}}
   \newcommand{\bdel}{\bar{\delta}}
\newcommand{\bQ}{\bar{Q}}
\newcommand{\bK}{\bar{K}}
\newcommand{\bq}{\bar{q}}

 \newcommand{\dphi}{\dot{\phi}}

 \newcommand{\lea}{{\leftarrow}}

            \newcommand{\pet}{{\cal P}}
\newcommand{\bapet}{\bar{\cal P}}

\newcommand{\ca}{{\cal C}}
\newcommand{\baca}{\bar{\cal C}}

            \newcommand{\beq}{\begin{quote}}
            \newcommand{\eq}{\end{quote}}
            \newcommand{\Om}{\Omega}
            \newcommand{\al}{\alpha}
            \newcommand{\ben}{\begin{enumerate}}
            \newcommand{\een}{\end{enumerate}}
            \newcommand{\bit}{\begin{itemize}}
            \newcommand{\ei}{\end{itemize}}
    	\newcommand{\nn}{\nonumber}
            \newcommand{\r}[1]{(\ref{e:#1})}
            \newcommand{\edfl}[1]{\Label{#1}\end{df}}

\newcommand{\vb}{{\cal h}}
\newcommand{\hb}{{\cal i}}

\newcommand{\ve}{{\varepsilon}}

\newcommand{\dif}{{\partial}}
\newcommand{\half}{\frac{1}{2}}

\begin{document}
\begin{titlepage}
\noindent
June 22, 2000\\

\vspace*{5 mm}
\vspace*{20mm}
\begin{center}{\LARGE\bf
On the quantum BRST structure of classical mechanics.}\end{center}
\vspace*{3 mm}
\begin{center}
\vspace*{3 mm}

\begin{center}Robert
Marnelius\footnote{E-mail: tferm@fy.chalmers.se}
 \\ \vspace*{7 mm} {\sl
Institute of Theoretical Physics\\ Chalmers University of Technology\\
G\"{o}teborg University\\
S-412 96  G\"{o}teborg, Sweden}\end{center}
\vspace*{25 mm}
\begin{abstract}
The BRST-antiBRST invariant path integral formulation of classical
mechanics of Gozzi
et al is generalized to pseudomechanics.   It is shown that projections to
physical
propagators may be obtained by BRST-antiBRST invariant boundary conditions. The
formulation is also viewed from  recent group theoretical results within
BRST-antiBRST invariant theories. A natural bracket expressed in terms of
BRST and
antiBRST charges in the extended formulation is shown to be equal to the Poisson
bracket. Several remarks on the operator formulation are made.
\end{abstract}\end{center}\end{titlepage}

\setcounter{page}{1}
\setcounter{equation}{0}
\section{Introduction.}
Gozzi {\em et al} have in a series of papers developed a path integral
formulation
of classical mechanics \cite{Goz}-\cite{GR}. More precisely they have given a
BRST-antiBRST quantum formulation of classical mechanics in which $\hbar$
is strictly
one. In this paper their treatment is further generalized and interpreted
from the
 BFV-formulation and viewed from  recent general
group theoretic results for BRST-antiBRST theories. In section 2 the path
integrals
for classical mechanics is generalized to pseudoclassical mechanics in
which some
coordinates are odd Grassmann numbers. In section 3 it is pointed out that
there is
also an operator BRST formulation of classical mechanics, and it is  shown
how the
ghost extended propagators are reduced to propagators in classical mechanics by
BRST-antiBRST invariant boundary conditions. In section 4 the Hamiltonian
flow is
viewed from  recent general group theoretical results for BRST-antiBRST
theories.
These results contain a new Poisson bracket, the Q-bracket, defined in
terms of the
BRST and antiBRST charges which is shown to yield the correct Poisson
bracket for
classical mechanics. The general group theoretical results also contain
extended group
transformations which here are given explicitly for the Hamiltonian flow.  In
section 5 some remarks on quantization are given. In two appendices a
further extended
BRST-antiBRST algebra is given and the properties of the Q-brackets are
displayed.

\setcounter{equation}{0}
\section{Path integrals for pseudoclassical mechanics.}
In \cite{GRT2} the path integral formulation of classical mechanics was
generalized
to classical mechanics in arbitrary  coordinates on a symplectic manifold.
Here we
further generalize this treatment to pseudoclassical mechanics in arbitrary
coordinates on a supersymplectic manifold. We consider real coordinates
$\phi^a$ with
Grassmann parities $\ve_a\equiv\ve(\phi^a)=0,1$. In terms of super or
graded Poisson
bracket (PB) we have
\be
&&\{\phi^a, \phi^b\}=\om^{ab}(\phi).
\e{1}
Hamilton's equations are
 \be
&&\dphi^a=\{\phi^a, H(\phi)\}=\om^{ab}\dif_bH(\phi),
\e{2}
where $H(\phi)$ is a real Hamiltonian. The dot denotes derivative with
respect to time
$t$. All equations are local in $t$. The coordinates $\phi^a$
are assumed to span the manifold which means that $\om^{ab}(\phi)$ has an
inverse
$\om_{ab}(\phi)$ satisfying the properties
\be
&&\om_{ab}\om^{bc}=\om^{cb}\om_{ba}=\del_a^c, \quad
\ve(\om_{ab})=\ve(\om^{ab})=\ve_a+\ve_b.
\e{3}
$\om_{ab}$ are coefficients of the basic two-form $\Om$:
\be
&&\Om= d\phi^b\wedge d\phi^a\,\om_{ab}=\om_{ab} d\phi^b\wedge d\phi^a
(-1)^{\ve_a+\ve_b}.
\e{4}
Since $\Om$ is closed ($d\Om=0$) $\om_{ab}$ must satisfy
\be
&&\dif_a\om_{bc}(-1)^{(\ve_a+1)\ve_c}+cycle(a,b,c)=0,
\e{5}
which is equivalent to
\be
&&\om^{ad}\dif_d\om^{bc}(-1)^{\ve_a\ve_c}+cycle(a,b,c)=0,
\e{6}
which are the jacobi identities of \r{1}. Note the symmetry properties
\be
&&\om^{ab}=-\om^{ba}(-1)^{\ve_a\ve_b},\quad\om_{ab}=\om_{ba}(-1)^{(\ve_a+1)(
\ve_b+1)}.
\e{7}

The derivation of the corresponding path integral formulation follows now
from the
treatment in
\cite{Goz}-\cite{GR}.  We may define the
propagator by \cite{GRT2}
\be
&&P( \phi, t;\phi_0, 0)=\del(\phi-\phi_{cl}(\phi_0,t)),
\e{8}
where $\phi_{cl}$ is the solution of \r{2}.  The crucial steps in the
derivation of
the path integral formulation are first to view the delta function as a
functional
delta function and then to rewrite it as follows:
\be
&&\del(\phi-\phi_{cl})=\del(\dphi^a-\om^{ab}\dif_bH(\phi))
\mbox{sdet}(\del_b^a\dif_t-\dif_b(\om^{ac}\dif_cH(\phi)),
\e{9}
where `sdet' is the superdeterminant or the Berezian. Rewriting the delta
function in
terms of integration over Lagrange multipliers $\la_a$ and expressing the
superdeterminant as integrations over ghost variables leads then to the
path integral
formulation
\be
&&P(\phi, t;\phi_0, 0)=\int D\la D\pet D\ca \exp{\{i\int_0^t L_{eff}\}},
\e{10}
where the effective Lagrangian is
\be
&&L_{eff}=\la_a\dphi^a+\pet_a\dot{\ca}^a-
\la_a\om^{ab}\dif_bH-\pet_a\ca^b\dif_b(\om^{ac}\dif_cH)(-1)^{\ve_a+\ve_b}.
\e{11}
Compared to a path integral in quantum mechanics, $\hbar$ is strictly one
in \r{10}.
In \r{10} we have introduced the Lagrange multipliers $\la_a$,
$\ve(\la_a)=\ve_a$,
and the ghosts $\ca^a$ and $\pet_a$ with Grassmann parities,
$\ve(\ca^a)=\ve(\pet_a)=\ve_a+1$. ($\pet_a$ corresponds to $i\bar{\ca}_a$ in the
notation of \cite{Goz}-\cite{GR}.) The reality properties are
\be
&&(\phi^a)^*=\phi^a,\quad(\la_a)^*=\la_a(-1)^{\ve_a},
\quad(\ca^a)^*=-\ca^a(-1)^{\ve_a},\quad(\pet_a)^*=\pet_a,
\e{111}
and $H^*=H$. These reality properties
imply that $L_{eff}$ is real. (Notice that they imply
$(\om^{ab})^*=-\om^{ba}=\om^{ab}(-1)^{\ve_a\ve_b}$ and
$(\dif_aB)^*=B^*\stackrel{\lea}{\dif_a}$.) Since
the Lagrangian
\r{11} is given in a standard phase space form  we may immediately
read off  the effective Hamiltonian,
\be
&&H_{eff}=\la_a\om^{ab}\dif_bH+\pet_a\ca^b\dif_b(\om^{ac}\dif_cH)(-1)^{\ve_a
+\ve_b},
\e{12}
and that $(\phi^a, \la_a)$ and $(\ca^a, \pet_a)$ are canonically conjugate
pairs. The
fundamental nonzero elements of the extended PB  are therefore
\be
&&\{\phi^a, \la_b\}=\del^a_b,\quad\{\ca^a, \pet_b\}=\del^a_b,
\e{13}
which are consistent with the reality properties \r{111}. (Notice that this
bracket is
different from \r{1} since $\{\phi^a, \phi^b\}=0$ here.) The resulting
equations of
motion are
\be
&&\dphi^a=\{\phi^a, H_{eff}\}=\om^{ab}\dif_bH,\nn\\
&&\dot{\la}_a=\{\la_a,
H_{eff}\}=-\la_b\dif_a(\om^{bc}\dif_cH)(-1)^{\ve_a}-
\pet_c\ca^b\dif_a\dif_b(\om^{cd}\dif_dH)(-1)^{\ve_a+\ve_b+\ve_c},\nn\\
&&\dot{\ca}^a=\{\ca^a,
H_{eff}\}=\ca^b\dif_b(\om^{ac}\dif_cH)(-1)^{\ve_a+\ve_b},\nn\\
&&\dot{\pet}_a=\{\pet_a,
H_{eff}\}=-\pet_b\dif_a(\om^{dc}\dif_cH)(-1)^{(\ve_b+1)\ve_a},
\e{14}
which also follows from the Lagrangian \r{11}.
Notice that the original equation for $\phi^a$ is retained. In fact this
equation cannot be derived from a Lagrangian in the original coordinates
$\phi^a$
unless the two-form $\Om$ \r{4} is exact (see appendix A).

The effective action $S_{eff}=\int dt L_{eff}(t)$ is invariant under the BRST
transformations ($\ve$ is here an odd constant)
\be
&&\del \phi^a=\ve\ca^a(-1)^{\ve_a},\quad \del \la_a=\del\ca^a=0,\nn\\
&&\del \pet_a=-\ve\la_a,
\e{15}
as well as under the antiBRST transformations
\be
&&\bdel \phi^a=\ve\pet_b\om^{ba},\quad
\bdel\la_a=-\ve\pet_b\dif_a\om^{bc}\la_c(-1)^{\ve_a+\ve_c+\ve_a\ve_b},\nn\\
&&\bdel\ca^a=\ve\la_b\om^{ba}(-1)^{\ve_a}+
\ve\pet_b\ca^c\dif_c\om^{ba}(-1)^{\ve_a+\ve_b+\ve_c},\nn\\
&&\bdel
\pet_a=\half\ve\pet_c\pet_b\dif_a\om^{bc}(-1)^{\ve_a(\ve_b+\ve_c+1)+\ve_c}.
\e{16}
These transformations are generated by the BRST and antiBRST charges
\be
&&Q=\ca^a\la_a,\quad
\bQ=-\pet_a\la_b\om^{ba}-\half\pet_a\pet_b\ca^c\dif_c\om^{ba}(-1)^{\ve_b+\ve_c}
\e{17}
in terms of the  extended PB \r{13}. One may easily show that
\be
&&\{Q, Q\}=\{Q, \bQ\}=\{\bQ, \bQ\}=0,
\e{18}
where the last equality requires the Jacobi identities \r{6} of the
original PB \r{1}.
Obviously there is a relation between the original PB \r{1} and the extended PB
\r{13} with the BRST and the antiBRST charges. This connection will be given in
section 4 and appendix B. That the BRST and antiBRST charges, $Q$ and $\bQ$, are
conserved follows from the fact that the effective Hamiltonian \r{12} may
be written
as follows
\be
&&H_{eff}=-\{Q, \{\bQ, H\}\}=\{\bQ, \{Q, H\}\}.
\e{19}
The symmetry algebra \r{18} may be further extended to the following algebra
(cf.\cite{Goz}-\cite{GR})\footnote{Such an algebra was first given for
Yang-Mills
theories in
\cite{NO}, and for general Lie group theories within the BFV-scheme in
\cite{SH}.}
\be
&&\{Q, Q\}=\{Q, \bQ\}=\{\bQ, \bQ\}=0,\nn\\
&&\{\bK, Q\}=\bQ,\quad \{K, \bQ\}=Q,\nn\\
&&\{\bK, K\}=Q_g,\quad\{\bK, \bQ\}=\{K, Q\}=0,\nn\\
&&\{Q, Q_g\}=Q,\quad\{\bQ, Q_g\}=-\bQ,\nn\\
&&\{K, Q_g\}=2K,\quad\{\bK, Q_g\}=-2\bK,
\e{20}
where
\be
&&Q_g\equiv\pet_a\ca^a
\e{21}
is the ghost charge, and
\be
&&K\equiv\ca^a\om_{ab}\ca^b(-1)^{\ve_a},\quad
\bK\equiv\pet_a\om^{ab}\pet_b.
\e{22}
Notice the different symmetry properties of $\om_{ab}$ and $\om^{ab}$
\r{7}. Only $K$
involves the inverse of $\om^{ab}$, \ie $K$ exists only if $\om^{ab}$ is
nondegenerate. Furthermore, the relation $\{K, \bQ\}=Q$ requires the Jacobi
identities in the form \r{5}. One may easily check that
\be
&&\{Q_g, H_{eff}\}=\{K, H_{eff}\}=\{\bK, H_{eff}\}=0.
\e{23}
The reality properties  from \r{111} are that $\bQ$ and $Q_g$ are real
while $Q$, $K$,
and $\bK$ are imaginary. ($H_{eff}$ is real.)

\setcounter{equation}{0}
\section{Interpretations}
The BRST-antiBRST formulation of classical mechanics given in the previous
section is
a quantum gauge theory in which $\hbar=1$. The BRST charge $Q$ has the standard
BFV-form \cite{BFV} which tells us that we have an abelian gauge theory
where $\la_a$
are the original gauge generators. They make the phase space coordinates
$\phi^a$
completely arbitrary. In fact, all variables are unphysical which means that the
theory has no physical quantum degrees of freedom at all, which of course is the
reason why it may describe classical mechanics. The BFV-form of the effective
Lagrangian \r{11} is
\be
&&L_{eff}=\la_a\dphi^a+\pet_a\dot{\ca}^a-\{Q, \psi\},
\e{201}
where the gauge fixing fermion $\psi$ is
\be
&&\psi=\{H, \bQ\}=\pet_a\om^{ab}\dif_bH.
\e{202}
BFV prescribes the form $\psi=\pet_a\chi^a$ where $\chi^a$ are  gauge fixing
variables to
$\la_a$. Here we have $\chi^a=\om^{ab}\dif_bH$ which is an unconventional
choice.
Consistency requires the matrix
\be
&&\{\la_a, \chi^b\}=-\dif_a(\om^{bc}\dif_cH)(-1)^{\ve_a}
\e{203}
to be invertible, a condition which may be compared to the regularity of the
superdeterminant in the basic derivation \r{9}. This condition requires the
original
Hamiltonian $H$ to be regular. If $H$ is not regular then even the original
theory is
a gauge theory or at least a constrained theory and the effective theory
given here
is no longer appropriate. Notice that $H$ is only introduced through the
choice of
gauge fixing, which means that the time evolution of classical mechanics is
determined by the choice of gauge fixing. Indeed from a Lagrangian point of
view the
entire Hamilton's equation \r{2} is introduced through a choice of gauge fixing
\cite{GR}.

It is possible to transform the above path integral formulation of classical
mechanics into an operator formulation. One has then to impose the canonical
commutation relations from \r{13}
\be
&&[\phi^a, \la_b]=i\del^a_b,\quad[\ca^a, \pet_b]=i\del^a_b,
\e{204}
where the commutators are graded commutators defined by
\be
&&[A,B]=AB-BA(-1)^{\ve_A\ve_B}.
\e{205}
$\hbar$ is of course strictly one. The corresponding operator charges,
$Q,\;\bQ,\; K,
\;\bK$ are still given by \r{17} and \r{22}. However, the ghost charge
operator $Q_g$
is given by (cf.\r{21})
\be
&&Q_g=\half\bigl(\pet_a\ca^a-\ca^a\pet_a(-1)^{\ve_a}\bigr).
\e{206}
$\bQ$ and $Q_g$ are hermitian while $Q$, $K$, and $\bK$ are antihermitian. Their
algebra is given by \r{20} with PB's replaced by commutators multiplied by
$(-i)$.
The effective Hamiltonian operator is
\be
&&H_{eff}=[Q, [\bQ,
H]]=-\dif_bH\om^{ba}\la_a(-1)^{\ve_a+\ve_b}-
\ca^{b}\pet_a\dif_b(\om^{ac}\dif_cH)(-1)^{\ve_a\ve_b},
\e{207}
which at the `classical' level is equal to \r{12}.

It is natural to define propagators by
\be
&&\vb\phi',\ca'|e^{-itH_{eff}}|\phi, \ca\hb=\int D\la D\pet D\ca
\exp{\{i\int_0^t
L_{eff}\}},
\e{208}
where $|\phi, \ca\hb$ are eigenstates to $\phi^a$ and $\ca^a$, and where
$L_{eff}$ is
the Lagrangian \r{11}. Such propagators were also considered in
\cite{Goz}-\cite{GR}
and were used to consider topological aspects in \cite{GR}.  Classical mechanics
should emerge from BRST-antiBRST invariant boundary conditions. However, one may
notice that the boundary conditions
$\ca^a=\ca^{\prime a}=0$ are BRST invariant, and
$\pet_a={\pet'}_a=0$ antiBRST invariant. Thus, in order to have both BRST
and antiBRST
invariant boundary conditions both these conditions should be imposed which is
impossible since  $\ca^a$ and $\pet_a$ are canonically conjugate variables.
On the
other hand it seems possible to impose the asymmetric boundary conditions
$\ca^a={\pet'}_a=0$ or
$\ca^{\prime a}=\pet_a=0$ in which case we obtain the physical propagators
($\pet_a=0$ is equal to an integration over $\ca^a$)
\be
P(\phi',t;\phi,0)=\int d\ca \vb\phi', \ca'=0|e^{-itH_{eff}}|\phi, \ca\hb,
\e{2081}
or
\be
P(\phi',t;\phi,0)=\int d\ca' \vb\phi', \ca'|e^{-itH_{eff}}|\phi, \ca=0\hb,
\e{2082}
both of which correspond to the original derivation in \r{9} and \r{10} read
backwards. These projections are peculiar since we impose BRST invariant
boundary
conditions on one side and antiBRST invariant conditions on the other.  Now
it is also
possible to impose strictly BRST-antiBRST invariant boundary conditions
provided the set
$\{\phi^a\}$ contains an even number of both odd and even coordinates which
actually  is what the standard BFV-prescription requires.
(In a recent paper
\cite{MS} projections of extended propagators to physical ones were studied in
details mainly within the standard BFV-formulation. The prescriptions given
there are
also applicable to the present case.) In this case we have to perform a
polarization, \ie a split in coordinates and momenta. We should therefore
first make a
transformation to Darboux coordinates. However, to simplify matters we
assume that
$\om^{ab}$ is constant in the following. We may then split the phase space
variables
into coordinates and momenta as follows
\be
&&\phi^a=(x^{\al}, p^{\al}),\quad
\om^{ab}=\left(\begin{array}{cc}0,&\eta^{\al\beta}\\
-\eta^{\al\beta},&0\end{array}\right),
\e{209}
where $a, b=1,\ldots,2n$ and $\al, \beta=1,\ldots,n$ and where
$\eta^{\al\beta}$ is a
constant symmetric metric
($\eta^{\al\beta}=\eta^{\beta\al}(-1)^{\ve_{\al}\ve_{\beta}}$).
   We split also the remaining variables
\be
&&\la_a=(\la_{\al}, \rho_{\al}),\quad\ca^a=(\ca^{\al}, \baca^{\al}), \quad
\pet_a=(\pet_{\al}, \bapet_{\al}).
\e{210}
The BRST and antiBRST charges \r{17} may then be written as follows
\be
&&Q=\ca^{\al}\la_{\al}+\baca^{\al}\rho_{\al},
\quad\bQ=-\bapet_{\al}\la_{\beta}\eta^{\beta\al}+\pet_{\al}\rho_{\beta}\eta^
{\beta\al}.
\e{211}
We are now naturally led to the following  two simple BRST-antiBRST
invariant states
\be
&&\rho_{\al}|\psi\hb_1=\bapet_{\al}|\psi\hb_1=\ca^{\al}|\psi\hb_1=0,\nn\\
&&\la_{\al}|\psi\hb_2=\pet_{\al}|\psi\hb_2=\baca^{\al}|\psi\hb_2=0.
\e{212}
If we define wave functions by $\psi(x,p,\ca,\baca)\equiv\vb x,p,\ca,\baca|\psi\hb$
where $|x,p,\ca,\baca\hb$ are the same eigenstates as in \r{208}, then the wave
function representation of the states $|\psi\hb_{1,2}$ in \r{212} are
$\psi_1=\del(\ca)\phi_1(x)$ and $\psi_2=\del(\baca)\phi_2(p)$, where
$\phi_{1,2}$ may
be gauge fixed to $\phi_1(x)=\del(x-x_{cl})$ and $\phi_2(p)=\del(p-p_{cl})$.
Notice that $\phi_1(x)$ and $\phi_2(p)$ are not Fourier transforms of each other
since $x^{\al}$ and $p^{\al}$ are not canonically conjugate variables here. The
point object may therefore be localized in both coordinate and momentum space, a
property which only is valid in classical mechanics. From the above results
we may now
directly derive the appropriate BRST-antiBRST invariant boundary conditions
for the
projections of the extended propagator \r{208} to physical ones: We have the
coordinate space propagator relevant for $\psi_1$ given by
\be
&&P(x', t;x, 0)=\int dp' dp d\baca' d\baca\vb x', p', \baca', \ca'=0|
e^{-itH_{eff}}|x, p,
\baca,
\ca=0\hb,
\e{213}
and the momentum space propagator  relevant for $\psi_2$ given by
\be
&&P(p', t; p, 0)=\int dx' dx d\ca' d\ca\vb x', p', \baca'=0, \ca'|
e^{-itH_{eff}}|x, p, \baca=0,
\ca\hb.
\e{214}

The above expressions are a little bit heuristic: Firstly, we should have used
eigenstates to the ghost momenta, $\pet_a$, since they are hermitian
operators (see
\r{111}). Then we should have used indefinite metric so that half of the
hermitian
operators have imaginary eigenvalues, and the operator $e^{-itH_{eff}}$
should be
replaced by the hermitian operator $e^{-tH_{eff}}$ \cite{MS}.

\setcounter{equation}{0}
\section{Group theory within the BRST-antiBRST framework}
In \cite{BM} general group theory within a BRST-antiBRST framework was
studied. This
was done at the quantum operator level. The formulas may however easily be
rewritten
in terms of classical functions. A group transformed function $A(\psi)$ where
$\psi^{\al}$ are the group parameters were first assumed to satisfy the Lie
equations
\be
&&{A}(\psi)\stackrel{\lea}{{\nabla}}_{\al}\equiv
{A}(\psi)\stackrel{\lea}{\dif_{\al}}
-\{{A}(\psi),
{Y}_{\al}(\psi)\}=0,
\e{301}
where $\dif_{\al}\equiv \dif/\dif\psi^{\al}$, and where the connections
$Y_{\al}$ were
assumed to be of the form
\be
&&{Y}_{\al}(\psi)=\{Q,\{\bQ, X_{\al}(\psi)\}\}.
\e{302}
$Q$ and $\bQ$ are the BRST-antiBRST charges respectively. The integrability
conditions for $X_{\al}$ following from \r{301} are
\be
&&X_{\al}\stackrel{\lea}{\dif_{\beta}}-X_{\beta}
\stackrel{\lea}{\dif_{\al}}(-1)^{\ve_{\al}\ve_{\beta}}-
\{X_{\al},
X_{\beta}\}_{Q}=\{X_{\al\beta}, Q\}+\{Y_{\al\beta}, \bQ\},
\e{303}
where the new Q-bracket is defined by
\be
&&\{A, B\}_{Q}\equiv \half\biggl(\bigl\{\{A, \bQ\}, \{Q, B\}\bigr\}-\bigl\{\{A,
Q\}, \{\bQ, B\}\bigr\}\biggr).
\e{304}
This bracket satisfies all properties of a graded PB except for the Jacobi
identities
and Leibniz' rule (see appendix B). It is nondegenerate on one-fourth of the
unphysical part of the extended supersymplectic manifold. In the present
case all
variables are unphysical and one easily finds that
\be
&&\{\phi^a, \phi^b\}_{Q}=\om^{ab}(\phi).
\e{305}
Thus, on the original supersymplectic manifold the Q-bracket \r{304} is
equal to the
original PB \r{2}. Only the original coordinates $\phi^a$ span the
Q-bracket (see
appendix B). (In \cite{GRT1} another multiPB expression involving $Q$, $K$
and $\bK$
was shown to yield the original bracket for
$\om^{ab}$ constant.) Further properties of the Q-bracket for the present
case are
 given in appendix B.

Consider now the Lie equations \r{301} and their integrability conditions
\r{303}.
For Lie group theories $X_{\al\beta}$ and $Y_{\al\beta}$ in \r{303} may be
chosen to
be zero. This is also the case in the present case since we have an abelian
gauge
theory. The group aspect is in fact only the Hamiltonian flow. The only group
parameter is time
$t$. The Lie equations \r{301} reduces therefore to Hamilton's equation
\be
&&\dot{A}(t)-\{A(t), H_{eff}\}=0.
\e{306}
For the choices $X_{\al}\equiv\phi^a$, $X_{\beta}\equiv H$ in \r{303}, the
integrability conditions reduce to the original Hamilton's equation
\be
&&\dphi^a=\{\phi^a, H\}_Q=\om^{ab}\dif_bH.
\e{307}
This shows the generality of the connections between \r{306} and \r{307}.

In \cite{BM} the integrability conditions \r{303} for $X_{\al}$ as well as
$X_{\al\beta}$, $Y_{\al\beta}$ etc were shown to be possible to embed into
a master
equation. A general solution of this master equation was then given. This
solution is
unfortunately not directly applicable to the present case. Firstly, the
construction
assumes that the group generators are connected to $\la_a$ in the BRST
charge and not
just the Hamiltonian flow. Secondly, the explicit construction is  made
within a manifest Sp(2) representation of the BRST and antiBRST charges using a
different ghost number representation in which both $Q$ and $\bQ$ have
ghost number
plus one \cite{BLT}. Only in the case when $\om^{ab}$ is constant are $Q$
and $\bQ$ in
a manifest Sp(2) form here. However, even in this case the appropriate
ghost numbers
are  impossible to determine without an explicit form of $H$ ($H$ has the
new ghost
number minus two).

Although we cannot demonstrate the details of the Sp(2)-construction of
\cite{BM} in
the present case, we may extract the final consequences of that paper. In
\cite{BM}
it was shown that one naturally arrives at an extended formulation in which the
number of group parameters are quadrupled to a supersymmetric set. In this
extended
framework there are then extended forms of the BRST-antiBRST charges. In
the present
case the time parameter $t$ is  replaced by $t,\;\rho,\;
\theta,\;\bar{\theta}$ where
$t,\;\rho$ are even and $\theta,\;\bar{\theta}$ odd. Furthermore, one has to
introduce conjugate momenta to these variables and define a further
extended PB. We
have the conjugate pairs $(t,\pi),$ $(\rho,\sigma)$, $(\theta, \xi)$ and
$(\bar{\theta}, \bar{\xi})$ satisfying
\be
&&\{t, \pi\}_e=\{\rho, \sigma\}_e=\{\theta, \xi\}_e=\{\bar{\theta},
\bar{\xi}\}_e=1.
\e{308}
On this extended super symplectic manifold we may introduce the extended
BRST and
antiBRST charges $\Delta$ and $\bar{\Delta}$ given by
\be
&&\Delta\equiv
Q+\theta\pi+\rho\bar{\xi},\quad\bar{\Delta}\equiv\bQ+\bar{\theta}\pi-\rho\xi.
\e{309}
We have also the following extended charges
\be
&&R\equiv K+\theta\bar{\xi},\quad
\bar{R}\equiv\bK+\bar{\theta}\xi,\quad\tilde{Q}_g\equiv
Q_g-\theta\xi+\bar{\theta}\bar{\xi}.
\e{310}
They satisfy the same algebra in the extended PB \r{308} as
$Q,\;\bQ,\;K,\;\bK$ and
$Q_g$ satisfy in \r{20}.

According to \cite{BM} we have an extended Hamiltonian flow defined by
\be
&&\tilde{A}(t,\theta,\bar{\theta},\rho)=\exp{\{Ad(F)\}}A\exp{\{-Ad(F)\}},
\e{311}
where
\be
&&F(t,\theta,\bar{\theta},\rho)\equiv-\{\Delta,\{\bar{\Delta},
tH\}_e\}_e=\nn\\&&=-t\{Q,\{\bQ, H\}\}+\theta\{\bQ, H\}-\bar{\theta}\{Q,
H\}+\rho H,
\e{312}
and where $Ad(F)$ is defined by
\be
&&Ad(F)\equiv\bigl(F\stackrel{\lea}{\dif_A}\bigr)\om^{AB}\dif_B,
\e{313}
where in turn $\dif_A$ are derivatives with respect to $\Phi^A$ which
represent  all
variables in the extended formulation ($\{\Phi^A, \Phi^A\}=\om^{AB}$).
Notice that the
previous Hamiltonian flow is obtained from
\r{311} if we impose
$\theta=\bar{\theta}=\rho=0$. In fact, any function $A$ of $\phi^a$
satisfies the
original Hamiltonian flow since $\tilde{A}(\phi, t)$ in \r{311} is
independent of
$\theta$,
$\bar{\theta}$ and $\rho$. The original equations \r{2} are not only
obtained from
\r{306}, but also  in terms of the Q-bracket in \r{307}, as well as  in
terms of the
extended bracket
\r{308}. One may also define a $\Delta$-bracket defined by
\be
&&\{A, B\}_{\Delta}\equiv \half\biggl(\bigl\{\{A, \bar{\Delta}\}_e, \{\Delta,
B\}_e\bigr\}_e-\bigl\{\{A, \Delta\}_e, \{\bar{\Delta}, B\}_e\bigr\}_e\biggr),
\e{314}
which is an extended $Q$-bracket (cf \r{304}). The $\Delta$-bracket is
spanned by
$\phi^a$ and $t$, $\sigma$. We have
\be
&&\{\phi^a, \phi^b\}_{\Delta}=\om^{ab}, \quad \{t, \sigma\}_{\Delta}=1.
\e{315}
Obviously (cf \r{307})
\be
&&\dphi^a=\{\phi^a, H\}_{\Delta}=\om^{ab}\dif_bH.
\e{316}

\setcounter{equation}{0}
\section{Remarks on quantization}
All formulas in the previous section may also be given in terms of operators and
commutators after a canonical quantization defined by \r{204}. There is then a
Q-commutator defined by
\be
&&[A, B]_{Q}\equiv \half\biggl(\bigl[[A, \bQ], [Q, B]\bigr]-\bigl[[A,
Q], [\bQ, B]\bigr]\biggr).
\e{401}
It follows that $[\phi^a, \phi^b]_{Q}$ satisfies all properties of a commutator.
However, since it is {\em not} a commutator, $\phi^a$ as operators are still
commuting. In fact, we have
\be
&&[A(\phi), B(\phi)]_Q=-i\{A(\phi),
B(\phi)\}_Q=-iA(\phi)\stackrel{\lea}{\dif_a}\om^{ab}\dif_b B(\phi).
\e{402}
Thus, even the operator formulation describes classical mechanics. It seems
therefore
as if the quantum BRST-antiBRST formulation of classical mechanics does not
contain
any new suggestion for quantization. (However, see \cite{AG} for an interesting
proposal.) A more conventional constraint formulation of classical
mechanics seems
therefore to be more hopeful for the quantization of mechanics on general
symplectic
manifolds \cite{FL,GL}.
\\
\\

\begin{appendix}
\setcounter{equation}{0}
\section{A further extended BRST-antiBRST algebra}
The original Hamiltonian equation \r{2}
cannot be derived from a Lagrangian written only in terms of $\phi^a$.
However, if
the two-form $\Om$ in \r{4} is exact then this is possible. An exact $\Om$ is
$\Om=2d\phi^a\wedge df_a(\phi)$ which implies
\be
&&\om_{ab}=\dif_af_b(\phi)+\dif_bf_a(\phi)(-1)^{(\ve_a+1)(\ve_b+1)}.
\e{a2}
The Lagrangian $L=f_a(\phi)\dphi^a-H(\phi)$ yields then the
equations $\om_{ab}\dphi^b=\dif_aH$ with $\om_{ab}$ given by \r{a2} which
are equal
to \r{2} due to \r{3}. In this case there is a natural extension of the
algebra \r{20}
(cf.\cite{Aux}). This extension involves the following three new generators
\be
&&q\equiv \ca^af_a(\phi),\quad\bq\equiv\pet_a\om^{ab}f_b(\phi)(-1)^{\ve_b},\nn\\
&&D\equiv\{Q,
\bq\}=\la_a\om^{ab}f_b(-1)^{\ve_b}+\pet_a\ca^b\dif_b(\om^{ac}f_c)(-1)^{\ve_a
+\ve_b+\ve_c}.
\e{a3}
The reality properties \r{111} requires $f_a^*=f_a(-1)^{\ve_a}$ since
$(\om_{ab})^*=-\om_{ba}$. The Lagrangian above is then real as well as the
generators
$\bq$ and $D$ while $q$ is imaginary.  The complete algebra is given by
\r{20} together with
\be
&&\{q,q\}=\{q,\bq\}=\{\bq, \bq\}=\{K,q\}=\{\bK, \bq\}=0,\nn\\
&&\{Q,q\}=-K,\quad\{\bQ,\bq\}=\bK,\quad\{K, \bq\}=q,\quad\{\bK, q\}=\bq,\nn\\
&&\{Q,\bq\}=D,\quad\{\bQ,q\}=-D-Q_g,\quad\{Q, D\}=\{\bq,D\}=0,\nn\\
&&\{q,D\}=-q,\quad\{\bQ,D\}=\bQ,\quad\{K, D\}=-K,\quad\{\bK,D\}=\bK,\nn\\
&&\{q,Q_g\}=q,\quad\{\bq,Q_g\}=-\bq,\quad\{D, Q_g\}=0.
\e{a4}
The new generators \r{a3} are, unfortunately, not conserved. This is
therefore not a symmetry algebra.\\ \\

\setcounter{equation}{0}
\section{Properties of the Q-bracket \r{304}}
Consider the Q-bracket \r{304}, \ie
\be
&&\{A, B\}_{Q}\equiv \half\biggl(\bigl\{\{A, \bQ\}, \{Q, B\}\bigr\}-\bigl\{\{A,
Q\}, \{\bQ, B\}\bigr\}\biggr).
\e{b1}
As stated in section 3 this bracket satisfies all properties of a graded PB
except for
the Jacobi identities and Leibniz' rule. In fact, it satisfies the
generalized Jacobi
identities \cite{BM}
\be
&&\{A,\{B, C\}_Q\}_Q(-1)^{\ve_A\ve_C}+
cycle(A,B,C)=\nn\\&&=
\biggl(2\{\{A, B\}_Q, \tilde{C}\}+\{\{\tilde{A}, B\}+\{A, \tilde{B}\},
C\}_Q\biggr)(-1)^{\ve_A\ve_C}+
cycle(A,B,C),
\e{b2}
where the tilde functions are defined by
\be
&&\tilde{f}\equiv\{Q, \{\bQ, f\}\}.
\e{b3}
Instead of Leibniz' rule the Q-bracket \r{b1} satisfies
\be
&&\{A, BC\}_Q-\{A, B\}_QC-B\{A, C\}_Q(-1)^{\ve_A\ve_B}=\nn\\
&&=-\{A, \{B, C\}_Q\}+\{\{A, B\},  C\}_Q+\{B,  \{A,C\}\}_Q(-1)^{\ve_A\ve_C}.
\e{b4}
Notice that Leibniz' rule is satisfied on a subset of `commuting' coordinates
provided the Q-bracket closes on this subset.

The BRST-antiBRST quantum theory for pseudoclassical mechanics leads to a
Q-bracket
with the following elements
\be
&&\{\phi^a, \phi^b\}_Q=\om^{ab}, \quad \{\phi^a,
\la_b\}_Q=-\half\dif_b\om^{ac}\la_c(-1)^{\ve_b+\ve_c+\ve_a\ve_b},\nn\\
&&\{\phi^a,
\ca^b\}_Q=\half\ca^c\dif_c\om^{ab}(-1)^{\ve_a+\ve_b+\ve_c},\quad\{\phi^a,
\pet_b\}_Q={3\over 2}\pet_c\dif_b\om^{ca}(-1)^{\ve_a+\ve_b+\ve_b(\ve_a+\ve_c)},\nn\\
&&\{\la_a, \la_b\}_Q=\{\la_a, \ca^b\}_Q=\{\ca^a, \ca^b\}_Q=\{\pet_a,
\pet_b\}_Q=0,\nn\\
&&\{\la_a,
\pet_b\}_Q=-\half\pet_c\dif_a\dif_b\om^{cd}\la_d(-1)^{\ve_b+\ve_d+\ve_c(\ve_
a+\ve_b)}
+\nn\\&&\quad+{1\over
4}\pet_d\pet_e\ca^c\dif_a\dif_b\dif_c\om^{ed}
(-1)^{\ve_c+\ve_e+(\ve_a+\ve_b)(\ve_c+\ve_d+\ve_e)},\nn\\&&\{\ca^a,
\pet_b\}_Q=-\half\dif_b\om^{ac}\la_c(-1)^{\ve_b+\ve_c+\ve_a\ve_b}+
\half\pet_d\ca^c\dif_c\dif_b\om^{da}(-1)^{\ve_b+\ve_c+\ve_d+\ve_b(\ve_a+\ve_d)}.
\e{b5}
The third line implies that the Q-bracket is trivial in the subsectors
$\{\la_a\}$,
$\{\ca^a\}$ and $\{\pet_a\}$. Only $\{\phi^a\}$ span the Q-bracket.
Leibniz' rule is
satisfied since the right-hand side of \r{b4} is zero for functions of
$\phi^a$. In
fact also the the Jacobi identities are satisfied. We have the following
basic tilde
operators:
\be
&&\tilde{\phi}^a=-\la_b\om^{ba}-\pet_b\ca^c\dif_c\om^{ba}(-1)^{\ve_b+\ve_c},
\nn\\
&&\tilde{\la}_a=\la_b\dif_a\om^{bc}\la_c(-1)^{\ve_a+\ve_c+\ve_a\ve_b}+
\pet_b\ca^c\dif_c\dif_a\om^{bd}\la_d
(-1)^{\ve_a+\ve_b+\ve_c+\ve_d+\ve_a\ve_b},\nn\\
&&\tilde{\ca}^a=-2\ca^c\dif_c\om^{ab}\la_b(-1)^{\ve_a+\ve_b+\ve_c}-
\pet_b\ca^c\ca^d\dif_d\dif_c\om^{ba}(-1)^{\ve_a+\ve_d},\nn\\
&&\tilde{\pet}_a=\half\pet_d\pet_b\ca^c\dif_c\dif_a\om^{bd}(-1)^{\ve_a+\ve_b
+\ve_a(\ve_b+\ve_d)}+
\pet_c\la_b\dif_a\om^{bc}(-1)^{\ve_a+\ve_a(\ve_b+\ve_c)}.
\e{b6}\\ \\

\end{appendix}

\end{document}